# Photos and Tags: A Method to Evaluate Privacy Behavior

Roba Darwish[1] and Kambiz Ghazinour[2]

[1] Colleges and Institutes Sector of Royal Commission Yanbu, Yanbu University College, Computer Science and Engineering Department, Yanbu, KSA
[2] Kent State University, Department of Computer Science, Kent, Ohio, USA
darwishr@rcyci.edu.sa , kghazino@kent.edu

**Abstract.** Online Social Networking Sites attracted a massive number of users over the past decade but also raised privacy concerns with the amount of personal information disclosed. Studies have shown that 25% of the users are not aware of privacy settings provided by these sites or do not know how to change them. This paper investigates an approach towards understanding users' privacy behavior on social media, e.g. Facebook, through studying faces, tags and photo privacy settings. It classifies users based on their privacy selections and proposes a system for monitoring and recommending stronger privacy settings. An application is developed, and our case study examines the effectiveness of our model.

## 1 Introduction

Social networking sites have become very important in our lives since their inception. They have revolutionized the world of communication, as they allow individuals to communicate with peers all over the world. However, the popularity of social networking sites presents the growing dilemma of preserving users' privacy.

Each social media site provides users with privacy settings. However, studies have shown [1, 2] that around 25% of social media users lean toward not to change, or are not by any means mindful of the service's privacy settings. As these services have become very popular over the past few years, more privacy risks are faced by users.

As of the first quarter of 2017, Facebook had 1.94 billion monthly active users, making it the largest social media service in terms of active users, and this number continues to grow [3]. Such prolific use was a prime motivation to investigate privacy issues and concerns on Facebook.

Users disclose their personal information, photos of themselves, their family and friends, ignoring the consequences of such behavior on their privacy.

This work inspects the use of faces and the presence of tags as a measurement tool to evaluate users' privacy behavior. Subsequently, new privacy categories are introduced and added to the existing categories defined by Alan Westin [4]. Moreover, we propose an application for monitoring and recommending better privacy settings for users. Machine learning techniques are also utilized in order to comprehend privacy settings of various users and suggest stronger settings for them.

This research contributes to the area of privacy in social media by:

• By using face detection, tags and their location on photos posted by individuals, a new method is proposed to measure behavior of privacy of the individuals.

• Based on the proposed new method, new privacy categories (besides existing Westin's privacy categories) are introduced.

• An application that monitors privacy settings for users and screens the privacy risks in their profiles. It then educates social media users of privacy-related issues, helping users to avoid them when using social media networks.

This recommender system provides benefits for both the user and the researcher. It suggests a better privacy setting and provides the user with the necessary steps on how to set up stronger privacy controls. It also has the ability to aid researchers by helping with data collection and the study of user demographics. Such information would then be used to design a social media network in which users are more aware of privacy settings, as well as future research in the field.

## 2   Background

### 2.1   Privacy on Social Media

Social networking sites have been developed significantly in recent times. Users aim to create many friends and connections. However, social media users are risking their privacy when utilizing such sites by disclosing a huge amount of personal information, images, messages and others, thus raising many privacy-related concerns.

By using social networking sites [6, 7] users expose themselves to many risks that significantly affect their privacy. Studies [6, 7] find that privacy could be attacked in a few ways if personal information was not provided reasonably and reliably.

Eecke and Truyens [7] discuss that there is an evidence that, once published, removing a "defamatory" or unwanted data from the Internet is nearly impossible. Moreover, they found that while different social network sites provide information protection tools to their users, the average user lacks an understanding of them, let alone properly using them. As mentioned earlier, most users tend not to change the default configuration set to make information publicly accessible. Subsequently, identity theft, stalking, physical harm, and other risks are increased.

### 2.2   Privacy on Facebook

Using Facebook, millions of individuals can create online accounts and share information with an enormous network that includes both friends and even strangers [8]. Facebook users usually disclose personal information, photos, and details about themselves and others. As the information disclosed increases, the risk of privacy violation increases as well [9]. Facebook content that is marked as "public" or with a little earth symbol  can be seen by almost everyone over the internet, including criminals, strangers and unwanted people [10]. They can discover many things about you, from what you like to your everyday activities and places you go. For instance, posting daily routes on Facebook or checking into nearby location of places you visit makes you an easy target for stalkers [11]. Other privacy risks result from offering your information to promoters. Moreover, this can be extended to creating a database gathered from individuals' profiles; this is done by third-parties who then sell the data to others. Passwords or even the whole database might also be stolen by intruders.

**2.2.1. Tagging.** Facebook allows users to add tags to their photos. Tags can be added to particular areas of your photo. For example, users can tag friends on their faces in a photo taken in a restaurant, tagging the restaurant with its name. According to the Facebook Help Center [12], the post that someone is tagged in it, can also be seen in that person's timeline. Hence, anyone who sees the posts can go to that person's profile to gain more information. To be more accurate, if the person tagged has a Facebook account, the tag is displayed as a hyperlink to that person's profile. If the tagged person does not have an account, then the tag is shown in plain text, not linking to any profile. So even if a user is concerned about privacy and has taken steps to protect his or her data by setting the privacy album visible to friends only, someone who is not a friend can still access that tagged photo and display the picture. Besides, friends of friends can also view that photo and find more information about the user. For example, it is easy to predict the age of a person through searching tagged Facebook photos and then checking posted birthday photos [13].

**2.2.2 Facebook Privacy Settings.** Facebook provides different privacy settings for users. Users can determine who can see their personal information. They can also set to whom their photo albums can be visible (e.g. Public, Friends, Friends of Friends, Custom, and Only Me). Table 1 shows privacy settings in Facebook.

**Table 1 –** Privacy settings available for Facebook

| Setting | Description |
| --- | --- |
| **Public** | When Facebook content is public, anyone can see it, even people you are not friends with, on or off Facebook. |
| **Friends** | Only those who are your friends on Facebook can access the content provided. |
| **Friends of Friends** | Your friends on Facebook and their friends can access the content provided. |
| **Custom** | Selective audience, specific people and networks can access the content provided. |
| **Only Me** | When you select the only me setting, only you can see the content you post. |

Unfortunately, most users do not know how to use these privacy settings [14]. They do not know how the settings work or simply do not have time to read them. This may account for most users leaving their privacy controls to the default setting, set to be visible to the public. According to Facebook, if the content is public, anyone over the internet can view it, including people who are not your friends and even those who do not have a Facebook account. In a survey of 200 Facebook users that aims to measure the disparity between the desired and actual privacy settings [14], researchers find that 36% of Facebook content remains shared with the default privacy setting. Likewise, they find that privacy settings match the users' desire only 37% of the time, and when incorrect, almost always expose content to a bigger number of users than anticipated.

When it comes to tagged photos, the default setting is that the tag is added to the Timeline and shared with the audience with no option available to the tagged person to review the content first [15]. This actually raises many privacy concerns.

It is worth noting that even if the selected setting for visibility is at "friends" level and the user has tagged someone in the post, the tagged person and their friends can view it. Similarly, the tagged person and their friends can see the post when the "only me" option is selected. In this work, the only tags considered are those found in images visible to the "public" and "friends of friends". Some of the user's information is always public on Facebook and users have no choice to change its setting. This information is listed in Table 2, along with the reason why Facebook makes it publicly visible. These reasons are provided by Facebook [12].

**Table 2 -** Public information on Facebook

| Public Information | Reason |
|---|---|
| **Name, Profile picture, cover photo** | Enables others to identify you. |
| **Gender** | Helps Facebook when referring to you (e.g. "add him as a friend"). |
| **Networks** | Enables others to find you (e.g. suggest friends who went to same school as yours). |
| **Username and user ID** | Displayed in the URL of any user account. |
| **Age** | Helps to give the user information that is appropriate for user's age. |
| **Language and country** | Enables Facebook to give the user a proper material. |

In this work we do not use face recognition methods to recognize celebrities and discard the photos that have pictures of people whose information is widely and publically available and posting them online would not necessarily violate their privacy (e.g. posting a picture of an ad in which a singer is promoting his/her concert).

## 3 RELATED WORK

### 3.1 Recommender Systems

Many researchers have recognized the problem of controlling privacy settings in social networking sites. They notice that people disclose a lot of information through social sites with little to no privacy context [16]. As a result, they propose recommendation systems in order to assist users to easily configure privacy settings. This section discusses some of recommender systems currently existing in the field.

In [17], the work presents a system which allows users to display the current privacy settings on their Facebook profile. It also detects possible privacy risks. The system monitors the privacy settings of user profiles and then recommends a setting for the user by using machine learning techniques to look at the similarity of the preferences chosen by the user who desires to set the privacy setting with other users who share common preferences.

Ghazinour et al. [5] are interested in the users' personal profiles, users' interests and users' privacy settings on photo albums to see whether they are visible to the public, friends, friends of friends, or on a custom setting. By observing how different users choose their settings on photo albums, the researchers classify them into one of three privacy categories. These privacy categories are specified by Alan Westin as Fundamentalists, Pragmatic and Unconcerned [4]. A detailed discussion is given in section B. Later, the Decision Tree is used to find different profile types. When it comes to recommendations, the K-Nearest Neighbor (KNN) algorithm was applied to predict the class of a new Facebook user by finding the similarities between their profile and others. Based on this, a recommendation is given by the system. The use of KNN makes this classifier like a collaborative filtering recommender system [18].

Ghazinour et al. [5] analyze users' data in order to understand their behavior in terms of how they choose their privacy settings. They found that most of the users shared information about their age, gender and education. However, when it comes to religion, political views and degree, users were more conservative. This study also shows a relationship between the users' interest and how they choose their own privacy settings. For instance, if the user's age is less than 21 years old, he/she usually belongs to the unconcerned category.

Mehatre and Chopde [16] propose a Privacy Policy Prediction system to assist users to compose privacy settings for their shared images. Their system relies on an image classification framework for image categories that might be linked with similar policies, and on a policy prediction algorithm to generate a policy for each newly uploaded image. This is done based on the user's social features. In summary, their proposed methodology is as follows: a) User uploads an image which has both objects and background. b) The object will be extracted and the background is suppressed to improve classification accuracy using foreground features. A saliency map, which is a kind of image segmentation, is used to help extracting image features. c) The extracted features are then compared with the database features of images. d) KNN is used to classify the class of the newly uploaded image. e) A policy comparison from the database is done by using a linear matching technique. f) Policy is accepted by the user.

Li et al. [8] present a trust-based privacy assignment system for social sharing that uses resources in social object networks. The presented system helps people select the privacy preference of the information being shared. This system, called the Personal Social Screen (PerCial), assists in assigning a privacy preference by automatically generating topic-sensitive communities users are interested in. It detects a two-level topic-sensitive community hierarchy before assigning a privacy preference to the users, depending on their personalized trust networks.

Ginjala et al. [19] introduce an intelligent semantics-based privacy configuration system (SPAC). This system automatically recommends privacy settings for social network users. SPAC uses machine learning techniques on both the privacy setting history and users' profiles to learn configuration patterns and then make predictions. In this system, semantics are integrated into the KNN classification to increase the accuracy of recommendations, such as semantics information in users' profiles.

Shripad and Vaidya [20] introduce a framework for handling trust in social networks based on a reputation mechanism. This mechanism works by capturing the implicit and explicit connections between the network users. It analyzes the semantics and dynamics of these connections. The system then provides personalized positive and negative user recommendations to another network user. The positive recommendations assist in connecting trustworthy users while the negatives alert users to not connect to untrustworthy users.

### 3.2 Westin's Privacy Categories

In our work users are classified into one of three privacy categories based on the setting chosen for their photo albums. These privacy categories are specified by Alan Westin, a researcher who conducted over 30 privacy surveys in the 1970s, as Fundamentalists, Pragmatic and Unconcerned. Privacy researchers around the world have used Westin's privacy index to measure attitudes and categorize people into these three groups. Descriptions of Westin's groups are provided below [21, 22].

Privacy Fundamentalist: Privacy Fundamentalists are highly concerned about their information. They are unwilling to provide any data or reveal information about themselves on websites when requested. People of this category tend to be worried about the precision of automated data and the additional uses made from it. They are agreeable to new laws that clearly explain an individual's rights and privacy policies. Privacy Fundamentalists form about 25% of the public.

Privacy Unconcerned: Privacy unconcerned people are willing to reveal and disclose any information upon request. They are the least protective of their privacy. Moreover, they do not favor expanded regulation to protect privacy. About 18% of the public are unconcerned.

Privacy Pragmatists: Privacy pragmatists are willing to disclose their information if they gain some benefits in return. Initially, they measure the potential advantages and disadvantages of sharing data with the organization. Next, they measure what protections are available for them and their trust level in the organization. Afterwards, they make a decision on whether or not to reveal their information to them and if revealing that information is actually worth it. 57% of people belong to this category.

## 4  Our Approach

Facebook currently owns the largest archive of personal photos. Thus, sharing and tagging images is actually built around real identities [23]. It is clear that the presence of faces and tags have a wide impact on the privacy of users. A photo showing a face of someone with a tag placed on the face could display the person's name, add a link to their profile for even more detailed information.

**4.1  Privacy Levels**

In this study, all conceivable instances of human faces and tags on user's photos are covered with the aim of categorizing them into various classes based on privacy categories explained in the next section. The following are different cases considered:

*4.1.1 CASE 1: No Faces, No Tags.* Photos that do not include faces and tags are considered to be the cases in which least privacy violation occurs. At the point when there is neither a face uncovering the identity of the individual nor a tag connecting directly to the user profile, no privacy violation is implied. However, cases exist that reveal information about the user whether a photo contains faces and tags or not. For instance, a user may post a photo of a landscape, in spite of the fact that there are neither faces nor tags, that photo may uncover other information about the user's present location.

*4.1.2 CASE 2: Some Faces, No Tags.* Having human faces in a photo may violate the privacy of an individual even if no tags are found. If the person is identified, information regarding the identity of the person is revealed and data with respect to the character of the individual is uncovered. This case is regarded as more privacy-revealing than Case 1.

*4.1.3 CASE 3: No Faces, Some Tags.* As mentioned earlier, tags link your profile to the post shared, making it easier for others to access your profile, view your information and discover more about you. This case is considered to disclose more private information than Case 2 (faces without tags) because through a profile, additional information can be uncovered, making it more valuable for the adversary than just knowing the person's face. For example, a friend of a friend can click on the tag link and view personal information such as location, hometown, relationships, and other information. Moreover, that user can view photos as long as they are visible to public or friends of friends. Keep it in mind that the profile picture, cover photos, and basic personal information are always visible to the public as clarified in Table 2.

*4.1.4 CASE 4: Some Faces, Some Tags.* The fourth case is when both faces and tags exist in a photo. In this case, whether the tag was placed on the face or not is examined. Answering this question will help in determining the privacy category of the user. Thus, if the tag is on the face, this facilitates in linking the tag name directly to the user, identify the user and get additional information from the user profile. If not, extra efforts were required in order to map the tag to the correct face in the photo which can sometimes be difficult. For instance, a photo that has a large number of faces and tags that are not placed on these faces makes it difficult for others to map the tag to the right face. However, this is still considered a violation of privacy.

This case incorporates three different cases. These cases are listed below:
  a) *No. of Tags < No. of Faces*
  b) *No. of Tags = No. of Faces*
  c) *No. of Tags > No. of Faces*

Those cases (with the same order in which they are mentioned above) and the tag not placed on the face, come next in privacy levels. The reasons for the order chosen are mentioned below with examples given to illustrate each case.

<u>**Tag Not on Face**</u>

***No. of Tags < No. of Faces.*** When the number of faces is greater than the number of tags, this might mean that there are faces existing on the photo that have no relation to any of the tags located in the photo. There can be faces of individuals with no names attached to it or links to their accounts. This refers to Case 2 (faces with no tags). Additionally, because these tags are not placed on the faces, there is minimal privacy risk involved compared to the case when the tags are positioned exactly on faces. To simplify this, it can be said that the tag does not directly relate the face to the person's name or the person's profile. For example, a photo which has three faces and one tag not placed on the person's face is less privacy-revealing than a photo having three faces and three tags placed on the faces. Figures 1 to 6 illustrate all the possible cases.

***No. of Tags = No. of Faces.*** When the number of tags is equal to the number of faces, assigning a tag to its related face would be much easier in case those tags belong to the existing faces on the photo. It would be much easier for adversary to map the person's face with the tag name and the attached user account. Figure 2 shows an example of having a photo with 3 faces and 3 tags with tags not placed on the faces.

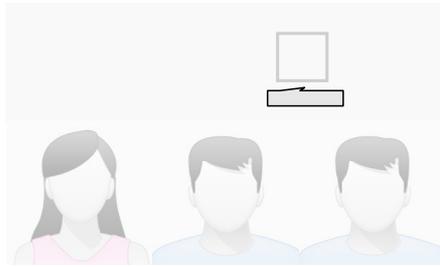

**Fig 1. A photo with less no. of tags than faces with tag not on face.**

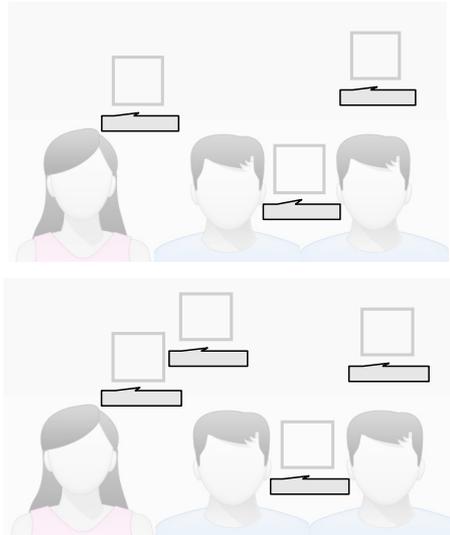

**Fig 3. A photo with greater no. of tags than faces with tag not on face.**

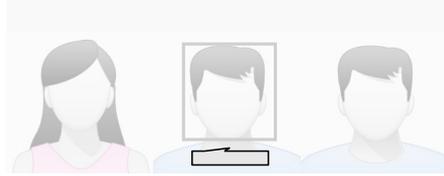

**Fig 4. A photo with less tags than faces with tag located on face.**

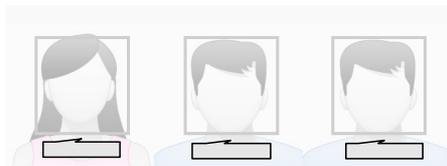

**Fig 5. A photo with equal number of tags and faces with tag located on face.**

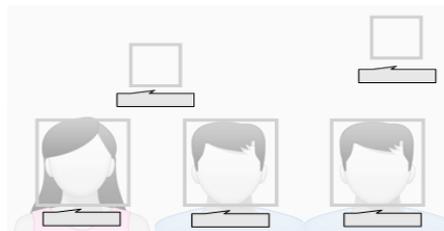

**Fig 6. A photo with more tags than faces with tag located on face.**

*No. of Tags > No. of Faces.* Some tags may not belong to any of the faces found in the picture. It is difficult to map tags to their corresponding faces, minimizing privacy violation. Most tags not placed on faces usually do not belong to the faces in the picture. However, in this case a user can discover more about other people from the tags available (as mentioned earlier, having more tags than faces violate more privacy).

For instance, if you have three faces and four tags in a photo as illustrated in Figure 3, the chance of finding personal information of others through their profile is higher.

### *Tag on Face*

*a) No. of Tags < No. of Faces.* As discussed earlier, having tags located on the person's face reveals a lot. By accessing a profile, adversary can easily identify a person's identity, appearance, and other revealing information such as personal information, photos, activities and much more.

Figure 4 shows three faces with one tag located on one of the faces. By looking at the picture, it is easy to figure out who the person is, as well as what he or she looks like. When it comes to other faces existing on the picture, privacy is also violated even if no tags are attached to them, albeit with a lower risk.

*b) No. of Tags = No. of Faces.* In this case, the tags directly relate the person's face with his/her name and profile account. This reveals much about people found in the photo with no effort needed to map the tags to the correct person. In other words, information about every person found in the photo is easily available. Figure 5 shows three faces with three tags located on them to illustrate the idea.

*c) No. of Tags > No. of Faces.* The most privacy revealing case compared to all other cases is having more tags than faces, with these tags located on the faces. In addition, since these tags are located on the face, the privacy risks are higher. Figure 6 shows a picture with three faces having tags located directly on them. Other tags also exist that reveal information about other people. Figure 7 concludes the cases as the higher up the case, the less private information can be revealed.

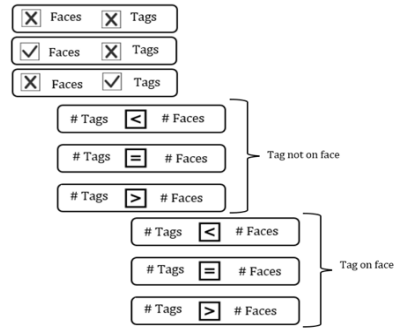

**Fig 7. Privacy levels of all cases covering faces and tags.**

### 4.2 User Classification, New Privacy Categories

As mentioned in Section 3.2, Westin categorizes people into three different groups: Fundamentalists, Pragmatics and Unconcerned. Ghazinour et al. [5] studied users' attitudes towards sharing their photo albums to determine the privacy category they belong to. The authors mention: "We also use photo albums since users treat them as a very personal and tangible type of personal identifiable information. Furthermore, it is one of the data items that Facebook allows us to check its privacy settings using the Facebook API functions" [5].

The authors use the following rules to determine the privacy category of each user for the profiling phase:

    If # of photos shared == 0 then:
        The user's privacy_category = **Fundamentalist**.
    Else if ratio of photos visible to Public or Friends of Friends < %50 then:
        The user's privacy_category = **Pragmatic**.
    Else:
        The user's privacy_category = **Unconcerned**.

In their work [5], Ghazinour et al. acknowledged that using the setting of photo albums as an indicator of unconcerned users may not be enough. They mentioned that having pictures of nature or art works in albums, which are set to be visible to public, do not imply any privacy violation which led us to take faces and tags in the pictures into account. In order to examine the model used in this study, as a first step, users were classified into one of five categories based on **faces** found in their photos. Three are the main categories derived by Westin. In addition, two new categories were introduced:

**Fundamentalist-Pragmatic (FP):** Fundamentalist leaning toward pragmatic. This group of people shares little data about themselves. The revealed information is less likely to violate any privacy. For instance, users who set their photo albums to be visible to public ensure that those photos shared do not contain any faces.

**Pragmatic-Unconcerned (PU):** pragmatic leaning toward unconcerned. This group of people share some data about themselves which also shows highly revealing information. An example of this is a user having only 10 photos in his profile, but with each photo containing many faces including the user and his friends.

The following rules are used to determine the privacy category of each user:

    If # of photos shared == 0 then:
        The user's privacy_category = **F**
    Else if all photos have faces then:
        The user's privacy_category = **U**
    Else if # of faces in all photos == 0 then:
        The user's privacy_category = **FP**
    Else if # faces < %50 of photos then:
        The user's privacy_category = **P**
    Else if # faces >= %50 of photos then:
        The user's privacy_category = **PU**

Second, users were classified based on both **faces** and **tags** into one of the seven categories using the following rules. New classes introduced here are P+, P and P- where P+ is more privacy preserving due to lack of tag use compared to P and P-.

    If # of photos shared == 0 then:
        The user's privacy_category = F
    Else if (# of faces == 0) AND (# of tags == 0) then:
        The user's privacy_category = **FP**
    Else if (# of faces != 0) AND (# of tags == 0) then:
        The user's privacy_category = **P+**
    Else if (# of faces == 0) AND (# of tags != 0) then:
        The user's privacy_category = **P**
    Else if (# of faces != 0) AND (# of tags != 0) then:
        If (# of tags < # of faces) Then:
            The user's privacy_category = **P-**
        Else if (# of tags == # of faces) then:
            The user's privacy_category = **PU**
        Else if (# of tags > # of faces) then:
            The user's privacy_category = **U**

The difference between having the tag placed on the face or not was acknowledged; for simplicity, when classifying users, they were treated equally in the above rules.

There are many other ways possible to label the instances such as sorting based on the number of tags, etc. However, this method was chosen in order to show the importance of combining faces and tags in studying privacy behavior. Interested researchers are invited to select their desired method for labeling.

### 4.3 Implementation

In this study, we present a Facebook application enables researchers to collect information from Facebook users, creating a dataset. The application is hosted on a virtual machine located in our lab. This system is designed as a Facebook application in order to collect information of user's personal profile, user's interests or likes and user's privacy settings on photo albums.

The Facebook app is written in JavaScript and PHP to access the Facebook user`s profile and settings. The system uses JavaScript SDK which provides a rich set of client-side functionality to access Facebook's server-side API calls.

***4.3.1. Face Detection.*** Microsoft Cognitive Services (formerly Project Oxford) is utilized to detect human faces in user images visible to either public or friends of friends. The API detects human faces in an image as input, returning face rectangles for where in the image the faces are in the output.

***4.3.2. Tag Location.*** Facebook API provides the following: a) The time the tag was created. b) Tagging person, representing the user who added the tag. c) X and Y coordinates in the photo where the tag is. d) Names of friends tagged in each photo.

In our application, getting the x and y coordinates of the tags was of interest so that the coordinates returned by Microsoft Face API for the face could be compared with the coordinates returned by Facebook API for the tags. This is done in order to determine if the tag is on the face or not. For educational purposes, the app also displays the tagged names to the user.

***4.3.3. System Database.*** User data collected through Facebook API is stored in a secure database. In this work, phpMyAdmin, one of the most well-known applications for MySQL databases administration, is used. In this database, a table was created to store personal information such as age, birthday, gender, hometown, location, political view, relationship status, religion, education, degrees, etc. Lookup tables were designed for maintaining data integrity in our database environment. For instance, if a user is entering his/her relationship into a data item, the User Data table containing the relationship item can reference a lookup table to verify that only one of the specified values is inserted. However, in tables such as location, the values are inserted in lookup tables in case they did not exist before; this is done by comparing the location ID with the IDs existing in the table.

***4.3.4 Users' interests.*** Such as music, movies, TV shows, and books they like are stored, as well as information related to user's photos, like the number of faces and tags in each image and whether the tag is placed on the face (if any). These information where collected to help us with the classification process in which users with the same interests would be grouped together.

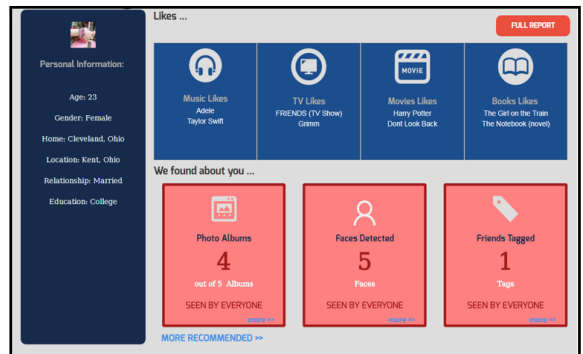

**Fig 8. The app interface where user's report displays personal information, interests and photo albums results.**

*4.3.5 Application Interface.* Users run the study's application and give permission to it to access their personal information disclosed on their profile. That user will then receive a report from the application that assists him/her in selecting a better privacy setting. This report summarizes the information provided to Facebook and, in a nutshell, notifies the user of his/her profile's privacy risks, suggesting a tighter setting. Figure 8 shows the report generated for the user where personal information, interests, publicly visible albums, number of faces detected, number of tags are displayed.

## 5 DATA COLLECTION AND ANALYSIS

### 5.1 Research Approvals

Before beginning data collection, the following approvals were obtained:

**Institutional Review Board (IRB).** Since the research involves collecting data from human subjects, it was necessary to submit the research proposal and supporting documents to IRB for approval. The IRB is responsible for protecting the rights and privacy of participants. Research conducted in this study using human participants was approved as a Level II/Expedited, category 7 project.

**Facebook approval.** Before releasing an application, Facebook needs to approve it. During the approval process, permissions required are submitted to Facebook through the App Dashboard. Some of the permissions submitted for review are: public_profile, user_birthday, user_education_history, user_hometown, user_location, user_relationships, user_religion_politics, user_likes and user_photos. Detailed instructions on how these permissions are used on the app, as well as screenshots, is also required.

### 5.2 Facebook User Data

In this study, it was of interest to collect the following user's data:

- *User's personal profile*: The attributes were collected from the user profile are: gender, birthday, education history, hometown, location, political view, relationship and religion.
- *User's interests*: This included books that users like to read, music they like to listen to, TV shows, and movies they like to watch.
- *User's privacy settings on photo albums*: Names, privacy settings, and cover photos of each album were collected.
- *User's photos:* URLs of photos in albums visible to public or friends of friends.
- *User's tags:* User's tagged friends in each photo of the publicly visible albums and friends of friends were collected.

The above parameters assisted in building a set of user data. When collecting user's photos, only recent photos are added. Using Facebook Graph API, the researchers were able to get the most recent ten photos of five albums posted within one year.

### 5.3 Data Harvesting and Participating Users

Users were recruited using Amazon Mechanical Turk (AMT). AMT is a service offered by Amazon that enables researchers to hire workers to perform human intelligence task. Workers were offered 1 dollar for running the app. 200 participants from 15 different countries with different educational backgrounds (median age 33; range 14 - 73) were asked to run the app at: [URL MASKED FOR BLIND REVIEW]

Participants were informed that the information they share will remain strictly confidential and will be used solely for the purposes of this research. Photos and images will not be stored. The data collected from them may be used verbatim in presentations and publications but the users themselves will not be identified. Results will be published in a pooled (aggregate) format. Moreover, the collected data is stored in a password-protected database at our lab where only the Principal Investigator and the co-Investigator have access to that database. Participants were informed that they needed to only run the app once with no need to answer any questions.

### 5.4 Data Preprocessing

Prior to running machine-learning algorithm for classification, it was necessary to prepare the collected data for further processing, including some preprocessing which was performed by the two Authors manually. A few issues faced and addressed were the following: Having multiple values referring to the same input. For instance, in the religion attribute, having values such as (Islam, islam, MUSLIM, Muslim, مسلم, مسلمة) referring to the same thing needed to be manually fixed so they all referred to one value "e.g. Muslim". This also applies for entries from different languages. For education attribute, values such as (Grad, Graduate School) needed to be changed to refer to the same value "e.g. Graduate". Changing location and hometown attributes from (city, state) or (city, country) format to (country) only limited the values we have (e.g. changing "Boston, MA" to "USA"). Entries such as (Christianity, Christian-Catholic) are also changed to their root value.

## 5.5 Data Analysis

After the preprocessing phase, data collected from participants are analyzed. When it comes to personal information such as gender, relationship, education, degree, location, hometown, religion and political view, users are comfortable disclosing some attributes but not others. The red color in the illustrated bar charts from Figure 9 and Figure 10 indicates the percentage of missing (not provided) values for the corresponding attribute. All participants provided their Gender with 0% of missing values. On the other hand, the highest percentage of missing values belonged to the Degree attribute. Table 3 shows the percentage of missing values for each attribute.

**Table 3. Percentage of values users did not disclose in this study**

| Attribute | % missing |
|---|---|
| Degree | 94% |
| Political view | 77% |
| Religion | 65% |
| Relationship | 41% |
| Hometown | 23% |
| Location | 20% |
| Education | 21% |
| Gender | 0% |
| Age | 0% |

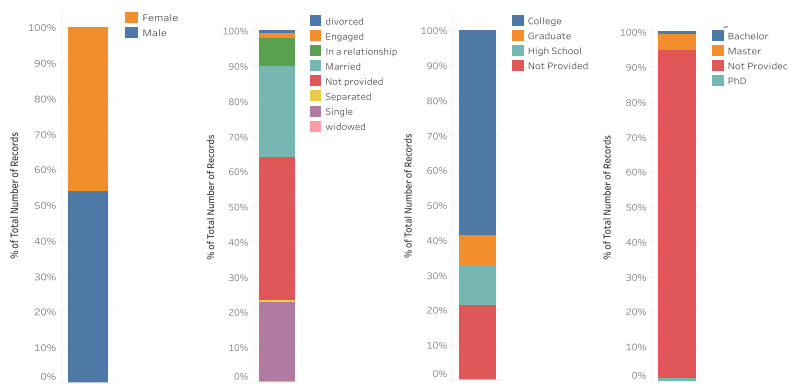

**Fig 9. Percentage of total number of records for gender, relationship, education and degree attributes (left to right).**

Concerning albums results collected from participants, the maximum number of albums is 25, while the average number visible to public is 10. Participants with an age range from 40 – 70 have the maximum number of 11 albums as the total number of albums they share. Moreover, participants with the age range of 14 – 29 have the maximum number of 50 faces as the total number of faces they share. When it comes to the total number of tags, the maximum is 23. The maximum number of faces and tags for participants with the age range: 30 – 70, are 28 and 10, respectively.

While both genders tend to share similar proportions in relation to the number of faces, albums, and public albums, there is a significant difference when it comes to the number of faces: males tend to share more faces than females. (see Figure 11)

Almost 32% of the users belong to the P+ class. This means that most of the users shared faces with no tags. 24% of the users belonged to the fundamentalist pragmatic category, thus sharing pictures with no faces or tags. 15% of participants were fundamentalists, while only 4% fell into the unconcerned category.

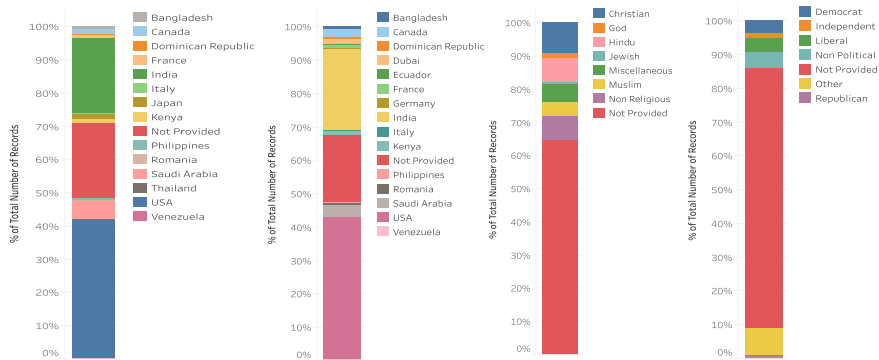

**Fig 10. Percentage of total number of records for hometown, location, religion and political view attributes (from left to right).**

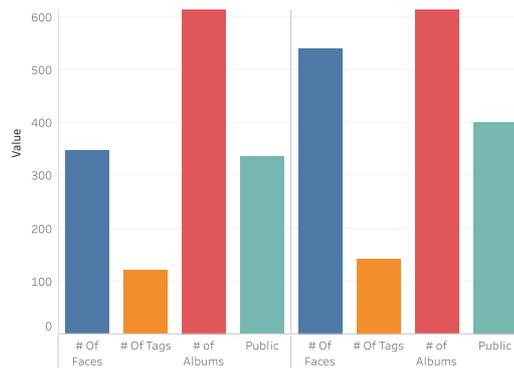

**Fig 11. Total no. of faces, tags, albums and public albums for both genders.**

## 5.6 User Classification

Using the rules discussed in Section IV. B, the Decision Tree algorithm in WEKA (J48) was used to categorize users into the following classes after labeling them:
- *3 Classes:* F, P and U.
- *5 Classes:* F, FP, P, PU and U.
- *7 Classes:* F, FP, P+, P, P-, PU and U.

When running Decision Tree algorithm on seven classes, the percentage of correctly classified instances is 87.85% using percentage split.

### 5.1 Recommender System

The KNN classifier is used to recommend a better privacy setting for the user. KNN ran on the 3 classes, 5 classes and then 7 classes of privacy behavior explained earlier. The goal is to compare the results in order to find the one with the highest prediction rate and thus the highest number of correctly classified instances. In all three different types of classification, participants are classified based on personal information disclosed on their profile. These are age, gender, hometown, location, relationship, religion, political views, education and degree. See [5] to understand how the recommender system works and suggests a less privacy revealing setting.

KNN is used with K= 4 for three different attributes. Those are education, location and relationship. A 66% Percentage split (66% training data and 34% test data with random split of dataset) is used as a test option. Results for different 3 classes, 5 classes and 7 classes of privacy behavior are shown in Tables 4, 5, and 6, respectively.

From the results shown above, the percentage of correctly classified instances for 7 classes of privacy behavior is not as strong as using 5 classes.

**Table 4. Three classes of privacy behavior for three attributes**

| Attribute | Education | Location | Relationship status |
|---|---|---|---|
| Correctly classified instances | 77.61% | 79.10% | 70.15% |

**Table 5. Five classes of privacy behavior for three attributes**

| Attribute | Education | Location | Relationship status |
|---|---|---|---|
| Correctly classified instances | 83.85% | 79.10% | 59.70% |

**Table 6. Seven classes of privacy behavior for three attributes**

| Attribute | Education | Location | Relationship status |
|---|---|---|---|
| Correctly classified instances | 80.59% | 83.58% | 77.61% |

### 5.2. Results and Main Findings

A study is conducted using the data collected from participants. The main results of the study are displayed below:

Users are comfortable disclosing their age and gender. They are more conservative when it came to sharing their location and hometown. More than 60% of participants are not willing to share their religions and political views. 94% of participants do not disclose their degree. Participants under the age of 30 reveal twice as many faces and tags in their posted photos than others. When the participants are classified into 7 categories, a large percentage of users are found to share many faces but no tags.

When recommending a better (less revealing) setting using KNN, the three and five Privacy classes give a better overall prediction rate than the seven classes.

### 6. CONCLUSION AND FUTURE WORK

The rapid growth of social networking sites has a negative impact on the privacy of the individual. Although privacy settings are available at these sites, they are not used to serve the user because the user is unaware of them, they are difficult to control, or the user simply does not care about them.

In this work, we introduced an application where privacy settings set by Facebook users are monitored and better settings are recommended. The application displays a report to the user where major privacy holes detected in their profile are shown. A new method for evaluating privacy issues by considering existence of faces and tags on publicly visible photos posted by Facebook users is introduced. Based on the new method, new privacy categories are introduced to Westin's privacy categories.

In this study, data of 200 participants from 15 different countries are collected. Participants are classified into three, five and seven privacy categories using a decision tree algorithm. Finally, KNN (with K=4) is used to recommend a tighter setting.

Limiting the number of photos (recent photos only) to evaluate for faces and tags has advantages such as reducing the expected waiting time to load the user report. However, this might affect the accuracy of measurement when classifying the user into various categories, since other photos might contain more faces and tags.

For future direction, processing all user's photos using a faster face detection method/algorithm is suggested. Furthermore, faces of famous people should be excluded from the face detection report. Another suggestion for future work is to analyze comments on posts since they reveal user information that may affect user privacy. For example, posting a beach image might not reveal anything, but a comment saying, "Relaxing in South Beach, Miami" can reveal where the user is her home is empty.

Finally, future work will concentrate on conducting further experiments considering a larger number of users. This is also needed in order to have a bigger training set, eventually resulting in better predictions. Moreover, in terms of interface design the aim is to display faces that have been detected in the user's photos; this can be used to warn the user and draw attention to the revealed people. Additionally, studying the impact of the application in terms of the amount of awareness raised is also necessary. It is of interest to monitor the main privacy settings that they prefer to change to both before and after running the application.